\documentstyle[epsbox,11pt]{article}

\newlength{\dinwidth}
\newlength{\dinmargin}
\def\slepton{\widetilde \ell}

\def\sneu{\widetilde \nu}

\def\sd{\widetilde{d}}

\def\st{\widetilde{t}}
\def\sb{\widetilde{b}}

\def\sz1{{\widetilde{Z}}_{1}}

\def\szk{{\widetilde{Z}}_{k}}
\def\swl{{\widetilde{W}}_{1}}

\def\swi{{\widetilde{W}}_{i}}
\def\swlp{{\widetilde{W}}_{1}^+}
\def\swlm{{\widetilde{W}}_{1}^-}

\def\msz1{m_{\sz1}}

\def\mswl{m_{\swl}}

\def\nle{{\stackrel{<}{\sim}}}
\def\nge{{\stackrel{>}{\sim}}}

\def\stl{\st_{1}}
\def\stls{\st^{*}_{1}}

\def\mstl{m_{\stl}}

\def\tht{\theta_{t}}

\def\rb{\slash\hspace{-7pt}R}

\def\lam{\lambda'_{131}}

\begin{document}
~~~\\
\vspace{10mm}
\begin{flushright}
ITP-SU-96/02  \\
hep-ph/9601338
\end{flushright}
\begin{center}
  \begin{Large}
   \begin{bf}
Is a high $P_T$ muon of the $e^+p \to \mu^+ X$ event observed at HERA \\
a signature of the stop? \\
   \end{bf}
  \end{Large}
  \vspace{5mm}
  \begin{large}
Tadashi Kon \\
  \end{large}
Faculty of Engineering, Seikei University, Tokyo 180, Japan \\
kon@ge.seikei.ac.jp\\
 \vspace{3mm}
 \begin{large}
    Tetsuro Kobayashi\\
  \end{large}
Faculty of Engineering, 
Fukui Institute of Technology, Fukui 910, Japan \\
koba@ge.seikei.ac.jp\\
  \vspace{3mm}
 \begin{large}  
    Shoichi Kitamura \\
 \end{large}
Tokyo Metropolitan College of Allied Medical Sciences, Tokyo 116, Japan\\
    kitamura@post.metro-ms.ac.jp\\
  \vspace{5mm}
\end{center}
\vskip50pt
\begin{quotation}
\noindent
\begin{center}
{\bf Abstract}
\end{center}
We investigate the $e^+p \to \mu^+ X$ event 
with high transverse momenta observed at HERA (H1) and show that this event 
could be interpreted as a signature of the single production of the scalar 
top quark in a supersymmetric model with $R$-parity breaking interactions. 
The event topology of the H1 event is rather characteristic and 
in fact it can be simulated by our specific scenario if 
we reasonably choose our model parameters to be 
({\romannumeral 1}) 
$m_{\sd, \sb, \sneu}$ $\nge$ $1$TeV [$0.8$TeV] for 
$\lam$ $=$ $0.1$ [$0.05$] and 
({\romannumeral 2}) 
$\mswl$ $\nle$ $150$GeV, $100$GeV $\nle$ $\mstl$ $\nle$ $200$GeV 
and $\lam$ $\nge$ $0.05$. 
\end{quotation}
\vfill\eject
%
%
%\section{\it Introduction}

Recently the H1 group reported that an event with high transverse momenta had 
been observed \cite{sgmu} at HERA. 
The total data sample analyzed at the H1 group corresponds to 
an accumulated luminosity of 3.2pb$^{-1}$ in positron (27.5GeV) -- 
proton (820GeV) collisions and of 0.8pb$^{-1}$ in electron -- proton 
collisions. 
The event is characterized by 
({\romannumeral 1})  a large transverse momentum of the single muon $\mu^+$,  
$P_T(\mu)$ $=$ $23.4\pm2.4^{+7}_{-5}$GeV, 
({\romannumeral 2})  a small $\delta$ ($\equiv \sum E (1-\cos\theta)$), 
$\delta$ $=$ $19.2\pm1.6^{+3.0}_{-2.1}$GeV, where $E$ and $\theta$ denote 
the energy and angle of any detected particle, 
({\romannumeral 3})  a large transverse momentum of the total hadronic 
system, $P_T$(hadron) $=$ $42.1\pm4.2$GeV   
and 
({\romannumeral 4})  a large missing transverse momentum 
 $P_T$(miss) $=$ $18.7\pm4.8^{+5}_{-7}$GeV. 

Some possible interpretations of the event have been given in ref.\cite{sgmu}; 
\begin{description}
\item[(A)] production of high $P_T$ jets 
\item[(B)] $W$ production and its leptonic decay 
\item[(C)] A flavour changing neutral currents (FCNC) or leptoquark production 
\end {description}
However, it does not seem to give reasonable explanation of the 
event characterized by the topology mentioned above \cite{sgmu}. 
As for the scenario {\bf A}, 
the probability for the event being due to the production of two high 
$P_T$ jets, where one jet shows the signature of a muon, is smaller than 
$10^{-3}$. 
As for the scenario {\bf B},
the $W$ production and its leptonic decay would give a rather small 
transverse momentum of the total hadronic system. 
In fact, the Monte Carlo calculation shows that for 
$P_T$(hadron) $>$ $40$GeV the cross section is reduced to 7fb. 
That is, with one event seen in 4pb$^{-1}$ we are left with a 3\% 
probability for this interpretation of the event. 
As for the scenario {\bf C},
expected events would have balanced $P_T$ and a value of 
$\delta$ $=$ $2E_e$. 
This is due to the fact that events originated from the FCNC or 
leptoquark production would 
show topologies like neutral current deep inelastic events, but with the 
final state positron replaced by a muon. 
The fact that less than 1\% of neutral current events show a value of 
$\delta$ $<$ $20$ GeV disfavours this interpretation.

In this letter, we propose a possible explanation of the single muon event 
in the framework of the minimal supersymmetric (SUSY) standard model 
(MSSM). 
We will show that the high $P_T$ muon could appear from the single 
scalar top quark (stop) production through an $R$-parity breaking coupling 
at HERA.

%\section{\it Previous work}

In the previous work \cite{stopsg}, we have already shown that 
one of the signals of the single stop production 
to be detected at HERA is characterized by the high $P_T$ 
 spectrum of muons. 
First, for the sake of convenience 
we will briefly summarize the basic idea by referring to our previous 
work \cite{stopsg}. 
The discussion is based on the MSSM with an $R$-parity breaking (RB)
interaction  
\begin{equation}
L=\lambda'_{131}\cos\tht (\stl {\bar{d}} P_L e + \stls \bar{e} P_R d),  
\label{stRb}
\end{equation}
where $\lambda'_{131}$ and $\theta_t$, respectively, denote 
the coupling strength and the mixing angle of the stops \cite{stop,HK}. 
Here $P_{L,R}$ read left and right  handed chiral projection operators.   
The interaction Lagrangian (\ref{stRb}) has been originated from the general RB 
superpotential  \cite{Barger}; 
\begin{equation}
W_{\rb}=\lambda_{ijk}\hat{L}_i \hat{L}_j \hat{E^c}_k 
+ \lambda'_{ijk}\hat{L}_i \hat{Q}_j \hat{D^c}_k + 
\lambda''_{ijk}\hat{U^c}_i \hat{D^c}_j \hat{D^c}_k, 
\label{RBW}
\end{equation}
where $i, j, k$ are generation indices. 
The first two terms violate the lepton number $L$ and the last term 
violates the baryon number $B$. 
If we want to explain such unresolved problems as 
({\romannumeral 1}) the cosmic baryon number violation, 
({\romannumeral 2}) the origin of the masses and the 
magnetic moments of neutrinos and 
({\romannumeral 3}) some interesting rare processes 
in terms of the $L$ and/or $B$ violation, 
the $R$-parity breaking terms must be incorporated in the MSSM.

The coupling Eq. (\ref{stRb}) 
will be most suitable for the $ep$ collider experiments at HERA 
because the stop will be produced in the $s$-channel in $e$-$q$ sub-processes 
\cite{stopsg,stoprb} 
\begin{equation}
ep \to {\widetilde{t}}_1 X.
\end{equation}
Note that the stop cannot couple to any neutrinos via $R$-breaking
interactions. 
This is a unique property of the stop which could be useful for us to
distinguish 
the stop from some leptoquarks. 
Production processes of the first and second generation squarks have been 
discussed in ref.\cite{heraRB}.

Babu and Mohapatra \cite{babu} have recently shown that the severe 
  constraint on a product 
  $\lambda'_{113} \lambda'_{131}$ $\nle$ $3\times 10^{-8}$ 
  comes from experimental data of the neutrinoless double $\beta$ decays. 
  Here, we will assume $\lambda'_{131} $ to be only non-zero 
  coupling parameter in what follows.   The upper bound on the strength of 
  the coupling  has been investigated through the low-energy experiments 
  \cite{Barger} and 
  the neutrino physics \cite{Enqvist}.   The most stringent bound 
$\lambda'_{131} 
  \stackrel{<}{\sim} 0.25$  comes from the atomic parity violation experiment 
  \cite{Barger}.

Next we examine the decay modes of the stop. 
In the MSSM, the stop 
lighter than the other squarks and gluino 
can decay into the various final states : 
\begin{eqnarray*}
\stl &\to& t\,\szk   \qquad\qquad\qquad\qquad\qquad\qquad({\rm a}) \\
 &\to& b\,\swi   \ \ \quad\qquad\qquad\qquad\qquad\qquad({\rm b})\\
 &\to& b\,\ell\,\sneu \qquad\qquad\qquad\qquad\qquad\qquad({\rm c})\\
 &\to& b\,\nu\,\slepton \qquad\qquad\qquad\qquad\qquad\qquad({\rm d})\\
 &\to& b\,W\,\szk \ \qquad\qquad\qquad\qquad\qquad \ ({\rm e})\\
 &\to& b\,f\,\overline{f}\,\szk \qquad\qquad\qquad\qquad\qquad
 \ ({\rm f})\\
 &\to& c\,\sz1 \quad\qquad\qquad\qquad\qquad\qquad \ ({\rm g})\\
 &\to& e\,d, \ \ \quad\qquad\qquad\qquad\qquad\qquad \ ({\rm h})
\end{eqnarray*}
where $\szk$ ($k=1\sim 4$), $\swi$($i=1,2$), $\sneu$ and $\slepton$,
respectively, 
denote 
the neutralino, the chargino, the sneutrino and the charged slepton. 
(a) $\sim$ (g) are the $R$-parity conserving decay modes, while (h) is 
only realized through the RB couplings (\ref{stRb}). 

If we consider the stop with mass small enough 
in the case of the $R$ conserving coupling,  
the first five decay modes (a) to (e) are kinematically 
forbidden due to the observed top mass $m_{t}$$\simeq$175 GeV \cite{top} 
as well as the model independent 
lower mass bounds for sparticles ; 
$m_{\swl}$ $\nge$ 45 GeV, $m_{\slepton}$ $\nge$ 45 GeV and 
$m_{\sneu}$ $\nge$ 40 GeV. 
So (f) and (g) survive. 
Hikasa and Kobayashi \cite{HK} have shown that 
the one-loop mode (g) $\stl\to c\sz1$ dominates over the 
four-body mode (f) $\stl\to bff'\sz1$. 
So we can safely conclude that such a light stop will decay into 
the charm quark jet plus the missing momentum taken away 
by the neutralino with  almost 100$\%$ branching ratio. 
On the other hand, 
if we consider the RB coupling $\lambda'_{131}$ $>$ $0.01$, 
which roughly corresponds to the coupling strength to be detectable at HERA, 
the decay modes (c) to (g) are negligible due to their large power of 
$\alpha$ arising from 
multiparticle final state or one loop contribution. 
Then only two body decay modes (a), (b) and (h) are left for our purpose. 

 We have found \cite{stopsg} that 
if the stop is heavy enough, i.e.,  
$m_{\widetilde{t}_1}>m_b+m_{\widetilde{W}_k}$ and the RB coupling is 
comparable with the gauge or Yukawa coupling 
$ \lambda' {2 \atop  {131}}/{4\pi}\stackrel{<}{\sim} \alpha,\alpha_t $ 
there is a wide range of parameters where 
$BR(\widetilde{t}_1 \rightarrow b\widetilde{W}_k)$ 
dominance over $BR(\widetilde{t}_1 \rightarrow ed)$ is assumed.   

In this case, 
 we should take into account of the process 
\begin{equation}
           ep         \rightarrow      b  \widetilde{W}_k X,    
\label{bc} 
\end{equation}
where the virtual contributions of the 
sneutrino with the same RB coupling constants $\lambda'_{131}$ 
have, of course, been considered.  
 The differential cross section is given by 
\begin{eqnarray}  
&&  \frac{d\sigma}{dxdQ^2}(ep \rightarrow b\widetilde{W}_k X)  = 
\frac{\alpha \lambda '^{2}_{131}}{16 \hat{s}^2\sin^2\theta_W} 
\Big[|V_{11}|^2  \frac{(\hat{u}-m^{2}_b)(\hat{u}-m^{2}_{\widetilde{W}_k})}
      {(\hat{u}-m^{2}_{\widetilde{\nu}})^2}                     \nonumber\\
 &&      + \frac{\cos^2 \theta _t \hat{s}}
{(\hat{s}-m^{2}_{\widetilde{t}_1})^2-m^{2}_{\widetilde{t}_1}
{\Gamma}  ^{2}_{\widetilde{t}_1}}\Bigl((|G_L|^2+|G_R|^2)
(\hat{s}-m^{2}_b -m^{2}_{\widetilde{W}_k})
-4m_b m_{\widetilde{W}_k}{\rm Re}(G_R G^{*}_L)\Bigr)          \nonumber\\
&&   - \frac{2\cos^2 \theta _t \hat{s}(\hat{s}-m^{2}_{\widetilde{t}_1})}
{\Bigl((\hat{s}-m^{2}_{\widetilde{t}_1})^2 +m^{2}_{\widetilde{t}_1}
{\Gamma}  ^{2}_{\widetilde{t}_1} \Bigr)(\hat{u}-{m^{2}_{\widetilde{\nu}}}  )}
{\rm Re}\Bigl(V^{*}_{11}(G_R\hat{u}+G_Lm_b 
{m_{\widetilde{W}_k}}     )\Bigr)  \Big],   
\end{eqnarray}
with $\hat{s}=xs, \hat{t}=-Q^2$ and 
\begin{eqnarray}
&&G_L   \equiv     - \frac{m_b U^{*}_{k2} \cos\theta_t} {\sqrt{2}m_W
\cos\beta},  \\  
&&G_R  \equiv  V_{k1} \cos\theta_t+  \frac{m_t  V_{k2} \sin\theta_t} 
{\sqrt{2}m_{W} \sin \beta}.
\end{eqnarray}
Here 
$V_{kl}$ and $U_{kl}$ 
stand for the chargino mixing angles \cite{Nilles} .    
The mixing angles as well as  masses  of the neutralinos  $m_{\widetilde{Z}_i}$ 
and the charginos  $m_{\widetilde{W}_k}$ are determined from the basic
parameters 
in the MSSM ($\mu ,  \tan \beta, M_2$). 
We can see that the $e^+$ beam is more efficient than the $e^-$ one to 
distinguish the stop signal from the SM background.   
 This can be understood from the fact that the $e^+$ collides with 
 valence $d$-quark in the proton, while the $e^-$ does only with 
 sea $\bar{d}$-quarks.   
It is expected that the detectable cross sections $\sigma
\stackrel{>}{\sim}0.1$ pb 
for heavy stop with mass ${m_{\widetilde{t}_1 }}\stackrel{<}{\sim} 250$ GeV for 
$e^+$ beams.   As far as $e^-$ beams are concerned 
$e^-p \rightarrow b\widetilde{W}_k X$ would be detectable for 
${m_{\widetilde{t}_1}} \stackrel{<}{\sim} 170$ GeV.   
In our model the LSP, the lightest neutralino $\widetilde{Z}_1$ 
possibly decays into $R$-even particles via only non-zero RB coupling 
$ \lambda'_{131} $.     
A typical decay chain will be  
\begin{eqnarray}
       ep \rightarrow b\widetilde{W}_1 X  \rightarrow
(b\ell\nu\widetilde{Z}_1) X  
       \rightarrow b(\ell\nu(bd\nu))X. 
\label{signature}
\end{eqnarray}
 
   Thus, a possible typical signature of the stop production 
   $ep \rightarrow  b\widetilde{W}_1  X$ 
   would be 
   $b$-jet+lepton+${\ooalign{\hfil/\hfil\crcr$P$}}_T$ in the case of 
   no LSP decay or 
  2$b$-jets+jet+lepton+ ${\ooalign{\hfil/\hfil\crcr$P$}}_T$  
  in the case of the LSP decay via RB coupling.   
  One of the signals to be detected at HERA is characterized by the high $P_T$ 
  spectrum of muons. 
  The lower $P_T$ cut certainly makes the event distinctive from 
  its background. 
  The cross section $\sigma (e^+p \to \stl X \to b\swl X)$ varies from $1$ to 
  10pb depending on mass of the stop in the range of 100 $\sim$ 150GeV.

%\section{\it Present analyses}
Now it is the position to present our calculation for 
some kinematical distributions 
in the process (\ref{signature}), which will also be compared to 
the experimental distributions of the H1 event \cite{sgmu}.  

First, the $P_T$($\mu$) distribution of the expected number of events is 
shown in Fig.1. 
In the calculation, we take a typical set of model parameters, 
($\mu$, $M_2$, $\tan\beta$, $m_t$, $\tht$, $\lam$) $=$ 
($-300$GeV, $50$GeV, $2$, $175$GeV, $1.0$rad, $0.1$) 
and the integrated luminosity $3.2$pb$^{-1}$. 
In this case we get the lighter chargino mass $\mswl$ $=$ $63$GeV and 
the lightest neutralino mass $\msz1$ $=$ $28$GeV. 
For simplicity, 
the branching ratio $BR(\widetilde{W}_1\rightarrow \nu\mu\widetilde{Z}_1)$ is 
 assumed to be $\frac{1}{9}$\cite{Schimert}.   
The dependence on the branching ratio of the 
chargino will be discussed later. 
We find in Fig.1 that rather heavy stop, $\mstl$ $=$ $100$ $\sim$ $120$, 
could give a high $P_T$($\mu$) event at the present integrated luminosity. 

We show the $\delta$ distribution together with the experimental data in Fig.2. 
The observed value of $\delta$ is significantly smaller than 
the allowed maximum value $2E_e$. 
Such $\delta$ could be obtained only when the LSP does not decay 
via the RB couplings in the detector. 
This constraint leads to following important consequences. 
We can quantitatively present the requirement as 
$c\gamma\tau_{\sz1}$ $\nge$ $1$m, which corresponds to 
$\Gamma_{\sz1}$ $\nle$ $10^{-7}$eV. 
By calculating the 3-body LSP decay width, we get severe constraints, 
$m_{\sd, \sb, \sneu}$ $\nge$ $1$TeV [$0.8$TeV] for 
$\lam$ $=$ $0.1$ [$0.05$]. 
Another one 
is that the total transverse hadronic momentum should be supplied by a 
$b$ quark at the first vertex in the process (\ref{signature}). 
In other words, we can take $P_T$($b$) $\simeq$ $P_T$(hadron) in 
our model calculation. 
Note, moreover, that large masses of sfermions justify our assumption on the 
chargino decay, $BR(\widetilde{W}_1\rightarrow \nu\mu\widetilde{Z}_1)$ $\simeq$ 
$\frac{1}{9}$. 
When sfermions are sufficiently heavy the dominant contribution to the 
decay matrix elements comes from the $W$-boson exchange diagram, 
which has apparently universal fermion couplings.

Shown in Fig.3 is the $P_T$($b$) ($\simeq$ $P_T$(hadron)) distribution. 
From the figure, we find that heavier stop 
is favourable to simulate the large transverse hadronic momentum 
$\sim$ $40$GeV. 
Note that the maximum value of $P_T$($b$) is determined from the simple 
kinematics of the two-body decay, 
\begin{equation}
E_b^{max}={\frac{1}{2\mstl}}(\mstl^2 + m_b^2 - \mswl^2).
\label{Ebmax} 
\end{equation}
Consequently, we can restrict possible regions in ($\mswl$, $\mstl$) parameter 
space in terms of the experimental constraint, 
$P_T$($b$) $\nge$ $P_T^{exp}$(hadron) $\simeq$ $40$GeV.

We present such a experimentally favourable region (shaded area) in Fig.4. 
Horizontal (dotted) lines correspond to one event with $P_T$($\mu$) $>$ 20GeV 
to be expected. 
It is seen from the figure that constraints have been set 
on the masses $\mswl$ and  $\mstl$, if we seriously take into account the 
present data.
For example, the chargino mass is smaller than about $100$GeV for 
$\lam$ $\nle$ $0.1$ under the condition of $P_T$($b$) $\nge$ $40$GeV. 
Moreover, $\lam$ should be larger than about $0.05$ and 
the stop mass should be larger than about $100$GeV 
because of the LEP bound on $\mswl$ $>$ $45$GeV. 
As has already been mentioned, low energy experiments set the constraint 
$\lambda'_{131} \stackrel{<}{\sim} 0.25$.

%\section{\it Concluding remarks}

We have investigated a possible scenario to explain the H1 single muon 
event by the single production of the stop with an $R$-parity breaking 
interaction in the framework of the MSSM. 
The event topology of the H1 event is rather characteristic and 
in fact they can be simulated by our specific scenario if 
we restrict arbitrary model parameters to some reasonable ranges. 
In order to work our scenario, we must have 
({\romannumeral 1}) 
$m_{\sd, \sb, \sneu}$ $\nge$ $1$TeV [$0.8$TeV] for 
$\lam$ $=$ $0.1$ [$0.05$], 
({\romannumeral 2}) 
$\mswl$ $\nle$ $150$GeV, $100$GeV $\nle$ $\mstl$ $\nle$ $200$GeV 
and $\lam$ $\nge$ $0.05$.

Our scenario would be confirmed or rejected at 
LEP2 or next linear colliders through the search for 
$e^+e^- \to \swlp\swlm$. 
Certainly, the discovery of the chargino could reveal us that the nature 
is supersymmetric but could give no information as to 
whether or not the nature does not respect the $R$-parity. 
To confirm it, we should seek for the stop with the $R$-parity 
breaking interaction, e.g., through a search for 
$e^+e^- \to ed\stl$. 
Needless to say, for our purpose it would be highly desirable to carry out 
the high luminosity run at HERA. 

%\begin{flushleft}
%{\Large{\bf Acknowledgements}}
%\end{flushleft}

\vfill\eject

\vfill\eject

{\Large{\bf Figure Captions}}
\begin{description}

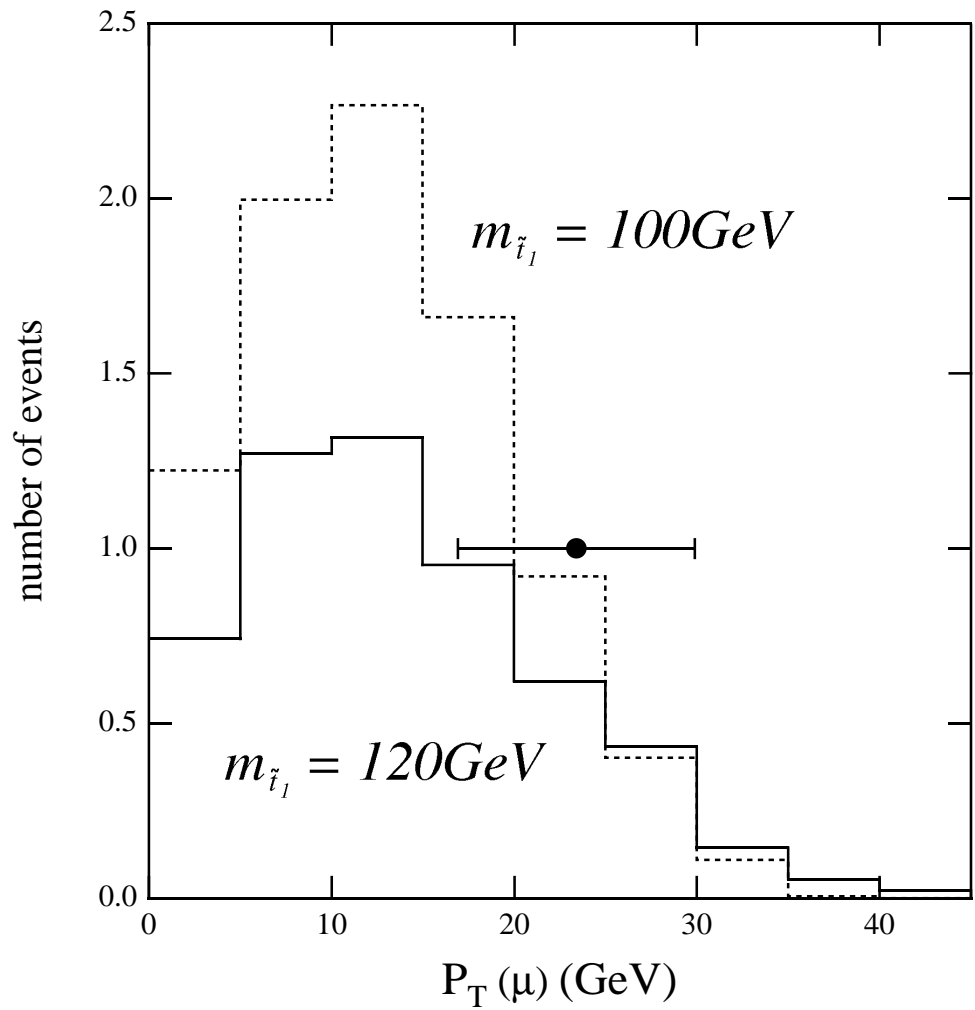
\item[{\bf Figure 1:}]
$P_T$($\mu$) distribution of the expected number of events 
together with the experimental data.
We take 
($\mu$, $M_2$, $\tan\beta$, $m_t$, $\tht$, $\lam$) $=$ 
($-300$GeV, $50$GeV, $2$, $175$GeV, $1.0$rad, $0.1$) 
and the integrated luminosity as $3.2$pb$^{-1}$. 
Solid line and dotted line respectively correspond to 
$\mstl$ =120 GeV and $\mstl$ =100 GeV.

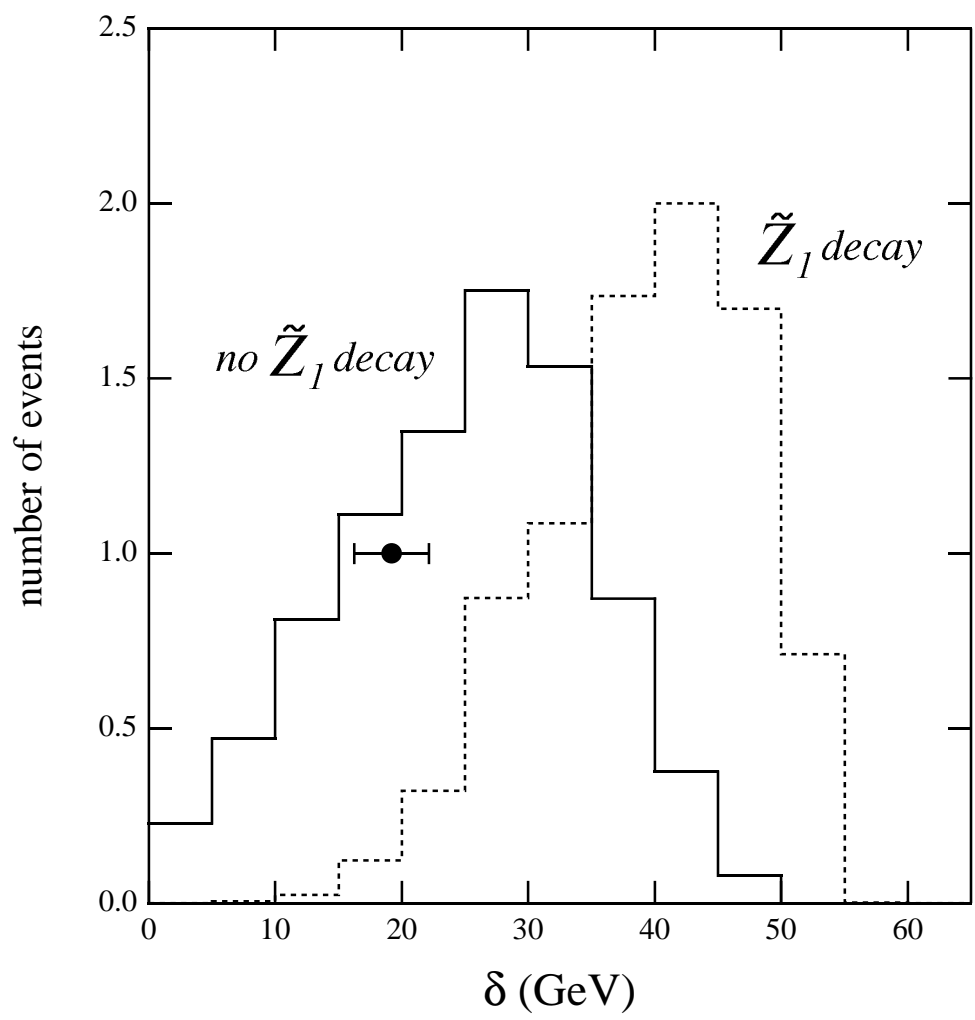
\item[{\bf Figure 2:}]
$\delta$ distribution of the expected number of events 
together with the experimental data.
We take $\mstl$ =100 GeV and the same parameters in Fig.1. 
Solid line and dotted line respectively correspond to 
no LSP decay and the LSP decay within the detector.

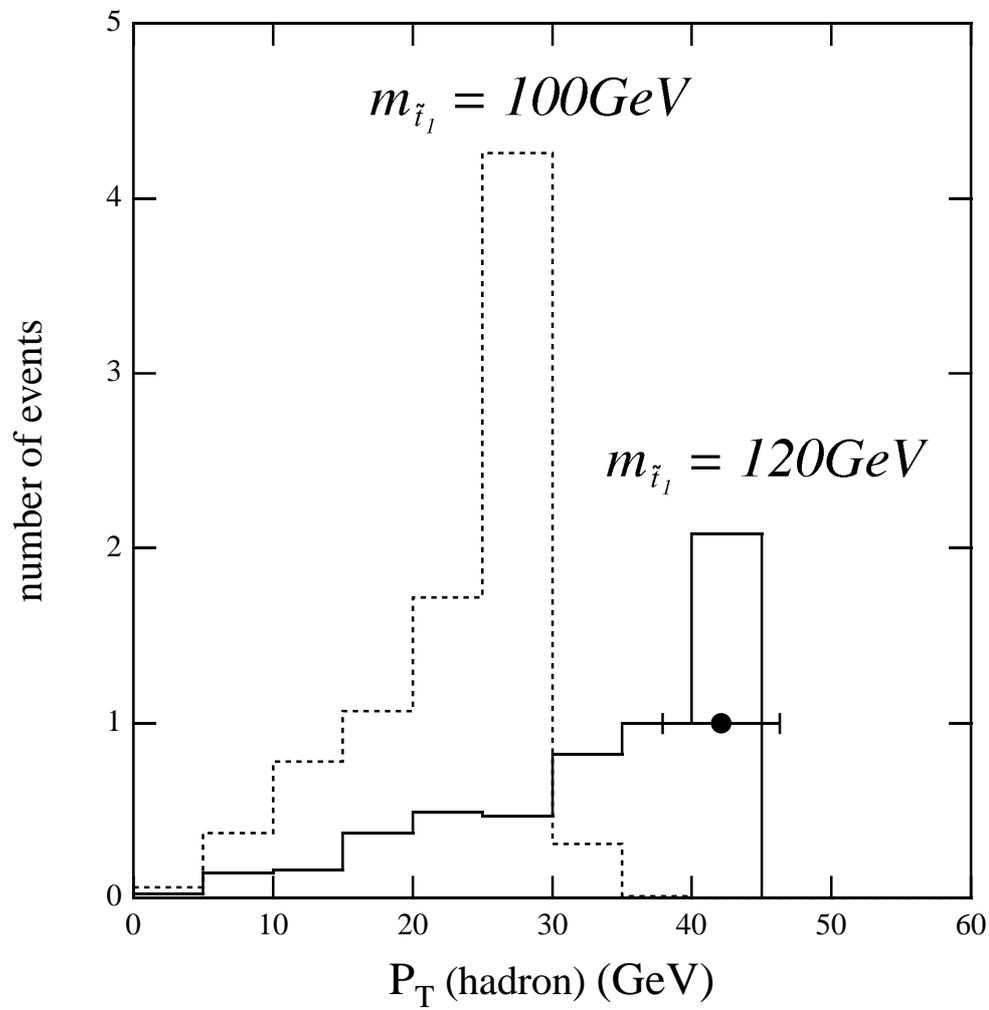
\item[{\bf Figure 3:}]
 $P_T$($b$) ($\simeq$ $P_T$(hadron)) distribution of the expected number 
 of events together with the experimental data.  
Parameters are the same in Fig.1. 

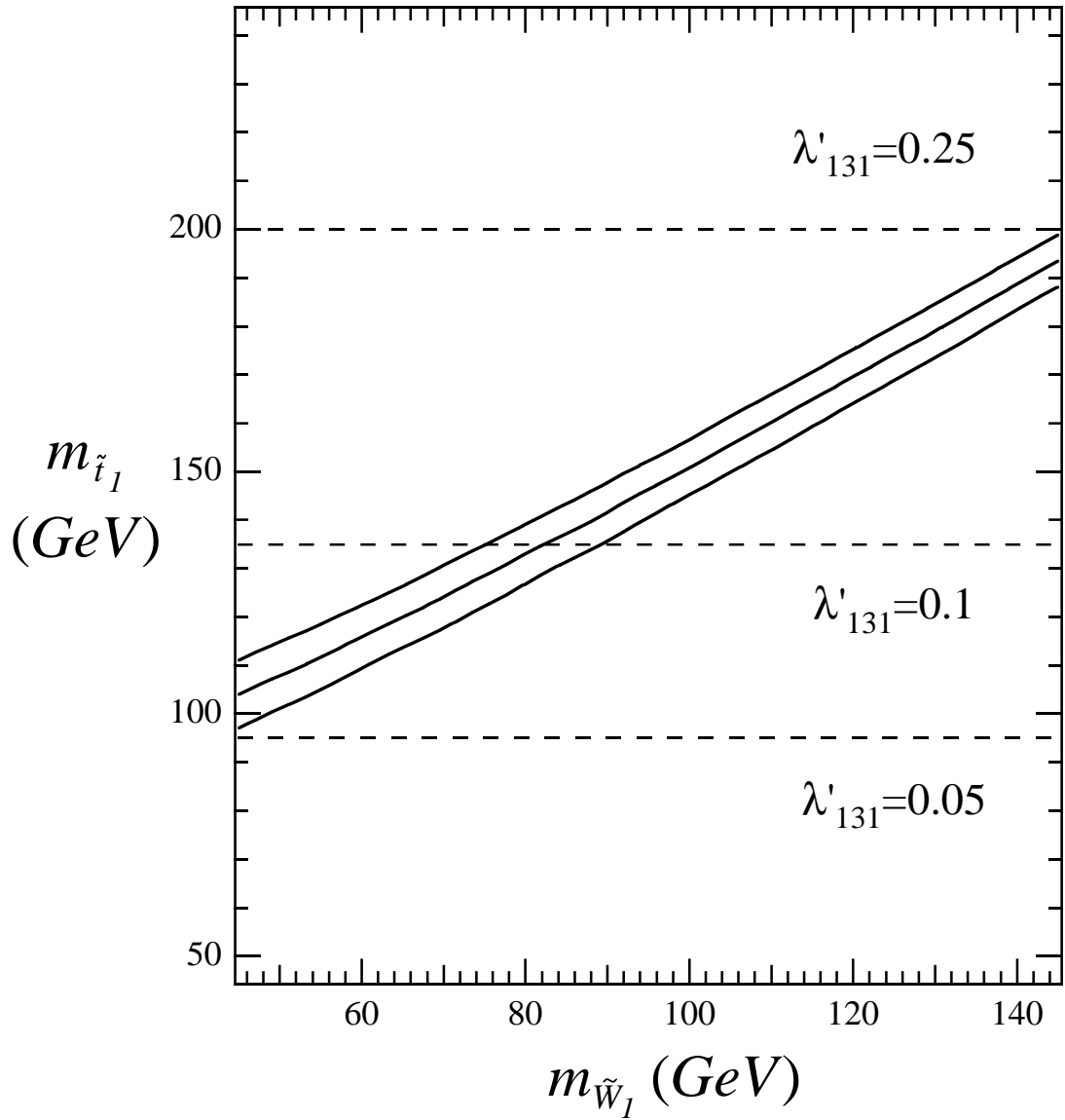
\item[{\bf Figure 4:}]
Favourable region (shaded area) in ($\mswl$, $\mstl$) parameter 
space by the experimental constraint, 
$P_T$($b$) $\simeq$ $P_T^{exp}$(hadron) $=$ $42.1\pm4.2$GeV. 
Horizontal (dotted) lines correspond to one event with $P_T$($\mu$) $>$ 20GeV 
to be expected. 
\end{description}

\vfill\eject


\begin{thebibliography}{99}
\bibitem{sgmu}
H1 Collab., DESY preprint, DESY 94-248
\bibitem{stopsg}
T. Kon, T. Kobayashi and S. Kitamura, {\it Phys. Lett.} {\bf B333}, 263
(1994) ; \\
T. Kobayashi,  S. Kitamura and T. Kon,  Seikei Univ. preprint, ITP-SU-95/03, 
hep-ph/9509294 
\bibitem{stop}
 J. Ellis and S. Rudaz, {\it Phys. Lett.} {\bf 128B}, 248  (1983); 
G. Altarelli and R. R\"uckl, {\it Phys. Lett.} {\bf 144B}, 126  (1984); 
I. Bigi and S. Rudaz, {\it Phys. Lett.} {\bf 153B}, 335  (1985)
\bibitem{HK}
  K. Hikasa and M. Kobayashi, {\it Phys. Rev.} {\bf D36}, 724 (1987)
\bibitem{Barger}
V. Barger, G. F. Giudice and T. Han, {\it Phys. Rev.} {\bf D40}, 2987 (1989)
\bibitem{stoprb}
  T. Kon and T. Kobayashi, {\it Phys. Lett.} {\bf B270}, 81 (1991)
\bibitem{heraRB}
 J. Butterworth and H. Dreiner, 
{\it Proc. of the HERA Workshop : "Physics at HERA"} 1991, 
eds. W. Buchm\"uller and G. Ingelman, 
Vol.2, p.1079 ; {\it Nucl. Phys.} {\bf B397}, 3 (1993), \\
J. L. Hewett, 
{\it "Research Directions for the Decade", 
Proc. of 1990 Summer Study on High Energy Physics, Snowmass, 1990}, 
ed. E. L. Berger, (World Scientific, Singapore, 1992), p.566 
\bibitem{babu}
K. S. Babu and R. N. Mohapatra, {\it Phys. Rev. Lett.} {\bf 75},  2276  (1995) 
\bibitem{Enqvist}
K. Enqvist, A. Masiero and A. Riotto, {\it Nucl. Phys.} {\bf   B373}, 95(1992)
\bibitem{top}
F. Abe et al. (CDF Collab.), {\it Phys. Rev. Lett.} {\bf 74}, 2626 (1995) ;
S. Abachi et al. (D0 Collab.), {\it Phys. Rev. Lett.} {\bf 74}, 2632 (1995)
\bibitem{Nilles}
  For reviews, see, 
H. Nilles, {\it Phys. Rep. } {\bf 110}, 1 (1984) ; 
H. Haber and G. Kane, {\it Phys. Rep. } {\bf 117},  75  (1985) 
\bibitem{Schimert}
T. Schimert, C. Burgess and X. Tata, {\it Phys. Rev.} {\bf  D32}, 707 (1985)
\end{thebibliography}
\end{document}